\documentclass[11pt,twoside]{article}
\usepackage{./asp2014}
\usepackage{color}
\usepackage{gensymb}
 \usepackage[koi8-r]{inputenc}
 \usepackage{color}
\aspSuppressVolSlug
\resetcounters
    
\renewcommand{\mag}[1]{^{\rm m}\!\!\!#1\,}
\newcommand{\Msun}{\ensuremath{\rm M_\odot}}
\newcommand{\Lsun}{\ensuremath{\rm L_\odot}}
\newcommand{\Rsun}{\ensuremath{\rm R_\odot}}
\newcommand{\kms}{\ensuremath{\mbox{km~s}^{-1}}}

\newcommand{\objeleven}{\rm Cyg~OB2~\#11~}

\begin{document}

\title{The study of massive stars with $50\,\Msun$ initial mass at different evolutionary stages}
\author{Olga Maryeva
\affil{Special Astrophysical Observatory of the Russian Academy of Sciences, Nizhnii Arkhyz, 369167, Russia; \email{olga.maryeva@gmail.com}}}

\paperauthor{O.\,V.\,Maryeva}{olga.maryeva@gmail.com}{ORCID_Or_Blank}{Special Astrophysical Observatory of the Russian Academy of Sciences}{Astrospectroscopy Laboratory}{Nizhnii Arkhyz}{Karachai-Cherkessian Republic}{369167}{Russia}

\begin{abstract}
 We will present results of studies of several massive stars at different evolutionary stages, but with similar values of the initial mass: O-supergiants belonging to association Cyg\,OB2, unique LBV/post-LBV -- Romano's star and two Wolf-Rayet stars -- WR156 and FSZ35. All these stars have similar initial mass of about $50~\Msun$. It allows us to consider them  a single star at different moments of life, and it  gives an opportunity to track changes in the physical parameters (such as effective temperature, luminosity, mass loss rate, wind velocity) and chemical abundances during the life of a massive star. It is important to test the current evolution  theories of such objects. 
\end{abstract}

\section{Introduction}
   
    According to modern concepts the evolution of massive stars  occurs as following: during hydrogen burning in the core  massive stars shift to the right on the Hertzschprung-Russell (H-R) diagram. OB-stars transform to red supergiants (RSG, $8\Msun<M_*<40\Msun$) or to luminous blue variables (LBVs, $40\Msun<M_*<60\Msun$) \citep{Meynet2011}. After that more massive stars ($M_*>30\Msun$) move again to the left part of H-R diagram and across the phase of Wolf-Rayet stars (WR), which is final stage before supernova (SN) explosion. For less massive stars ($8\Msun<M_*<30\Msun$) the progenitors of SN are directly RSG.

    Numerical modeling of the evolution of massive stars started in 90s. Due to a significant increase in computer processing power, modern codes for the calculation of stellar evolution have become reliable tools for astrophysical research -- for estimation of stellar ages, initial mass, chemical composition, etc. Modern stellar evolution codes generally describe well the locations of observed variety of the massive stars -- as well as RSG, LBV and WR -- on H-R diagram.

    However there are still a number of contentious issues  in the theory of massive stars' evolution. For further development of the theory it is necessary to compare theoretical predictions with observations, with the parameters of actually observed massive stars. One of the important tasks of modern stellar astrophysics is to determine the  stars' parameters on short stages of evolution such as LBVs, yellow hypergiants, blue supergiants, WR stars, and during the transitions between them.
    
    Here we present brief description of an evolution of massive star with 50-60\,$\Msun$ mass by means of studying various objects with such masses on different stages of their lives.

\section{Stars with 50-60\,$\Msun$ initial mass}
\subsection*{O supergiants}

   For consideration stage of O-supergiant was selected the stars from the association Cyg\,OB2 --  \#7 (${\rm O3If_*}$) and \#11 (${\rm O5.5Ifc}$)\footnote{Results was published in \citet{me2013} and \citet{me2014}}.

   Cyg\,OB2\,\#7 is one of the hottest stars in our Galaxy. As result of our modeling we found that in the spectrum of the star there are some lines whose are not described within the framework of single model with simple velocity law. For fitting strong emission line $H\alpha$ and absorption NIV and CIV lines is necessary or to involve considerations that wind is nonspherical, or to use more complicated velocity law. 
   

%

Cyg\,OB2\,\#11 is classified as ${\rm Ofc}$ supergiant \citep{WalbornOIfc}. Our results suggest that its physical parameters are close to the ones of ``normal'' O-supergiants with no strong carbon emission. In \objeleven the nitrogen abundance is lower than the one for other ``normal'' O stars, while the carbon abundance is nearly solar.
Therefore we may conclude that the mixing in atmospheres of Ofc stars that transports the products of CNO cycle from the core towards the stellar surface is somewhat damped.

\vspace{-5mm}
\articlefigure{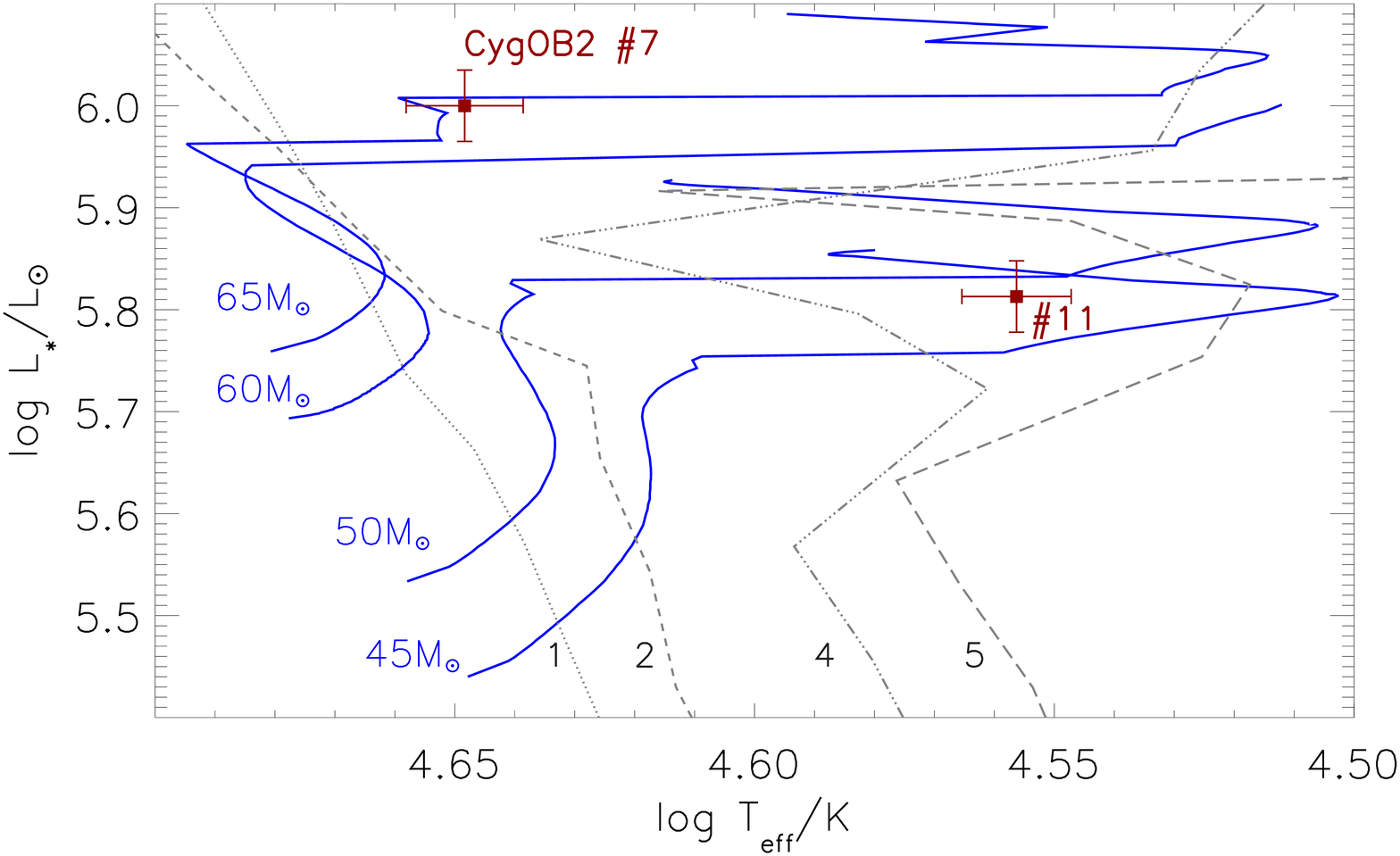}{fig_Ostar}{The locations of Cyg\,OB2\,\#7 and \#11 \ in the H-R diagram.
         The evolutionary tracks and the stellar isochrones are taken from the Geneva library.}
\vspace{-5mm}
\subsection*{Romano's star}

   Romano's star  (M33/V532 or GR290) is a famous variable star, which is now classified as LBV/post-LBV star and shows late-WN spectrum, and is very important for our understanding of evolution of massive stars in general. 
 
   Our analysis of nine  most representative spectra, obtained between October 2002 and December 2014, when the star displayed an ample range of variation in visual luminosity, shows that the bolometric luminosity of GR\,290 is variable, it is higher during the phases of greater optical brightness. It confirms the hypothesis from earlier works by \citet{polcaro10} and \citet{me2012pasha}. We also found that the structure of the stellar wind significantly changes, being much denser and slower during the eruption in 2005, while during the minimum of brightness the wind structure is fairly similar to the one of typical WN8h stars. 
    
   Figure~\ref{fig_HR} shows the recent path of GR\,290 in the H-R diagram  during two successive luminosity cycles. GR\,290 sits on the evolutionary track of a $\simeq$50~$\Msun$ star.  In maximum of brightness ({\it V}=$17\mag{\,}$, Feb.~2005)  the star is located on the  LBV minimum instability strip and it moves to  WR region in the minimum of brightness. GR\,290 is the first star which demonstrated the transition from the instability strip to WR region \citep{meBalticAstronomy,Polcaro2016}. The high effective temperature and WNL-type spectrum place the star after the low temperature loop of the evolutionary tracks which is thought to be occupied by the LBV stars. Estimated chemical abundances for atmosphere of GR\,290 show that the star is  younger than  WN8h stars \citep{Polcaro2016}.  Combining the results of numerical modeling with data of photometric and spectral monitoring we may conclude that we observed GR\,290 in very rare evolutionary phase -- post-LBV. 

 \vspace{-5mm}
\articlefigure{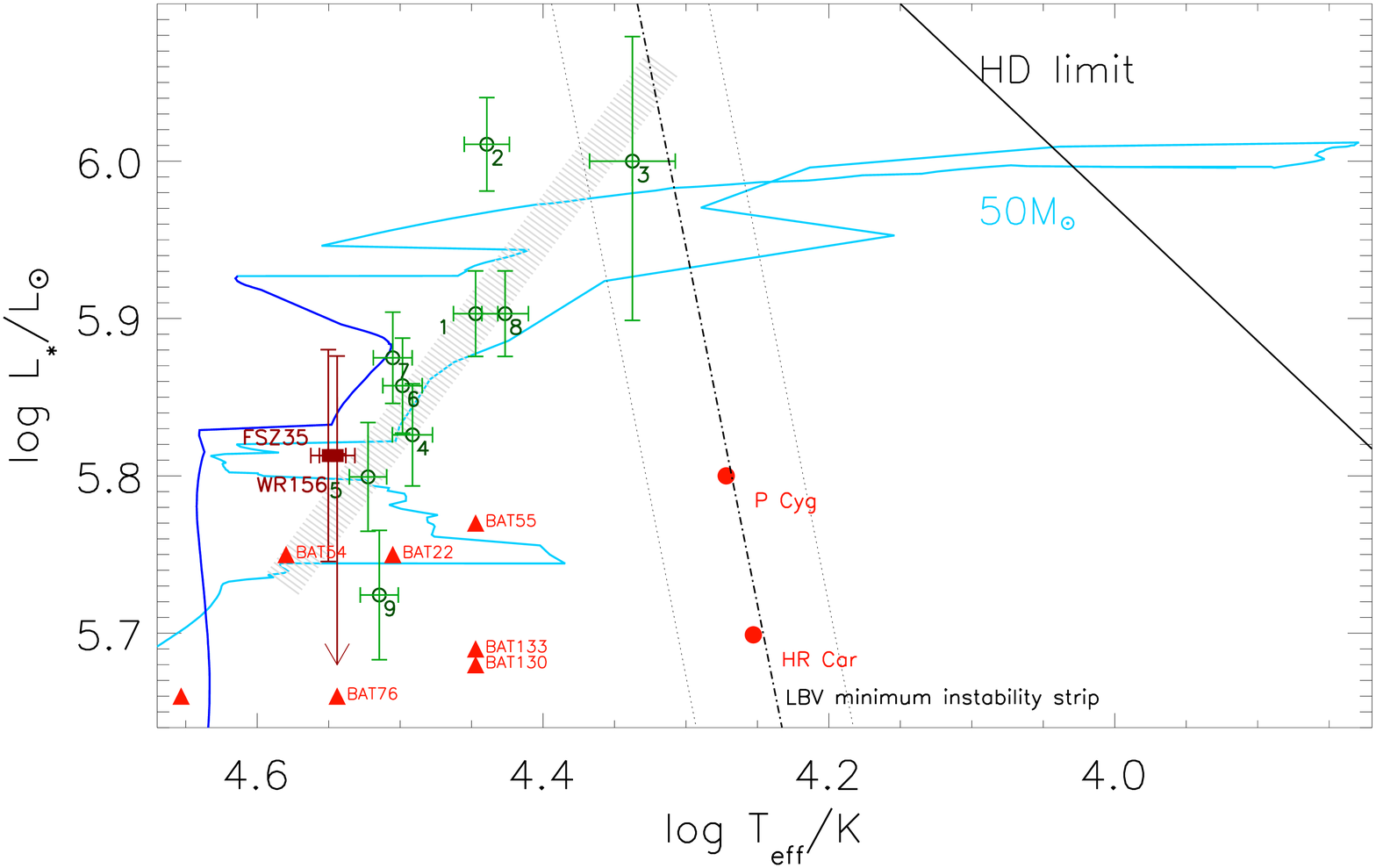}{fig_HR}{Position of GR\,290 in the H-R diagram in 2002 -- 2014 years. HD limit is Humphreys-Davidson limit \citep{HumphreysDavidson}. Numbers correspond to the dates of observations: 1 -- Oct 2002, 2 -- Feb 2003, 3 -- Jan 2005, 4 -- Sep 2006, 5 -- Oct 2007, 6 -- Dec 2008, 7 -- Oct 2009, 8 -- Dec 2010, 9 -- Aug 2014. Filled squares mark the positions of FSZ35 and WR156.}
 \vspace{-5mm}
   
 \subsection*{Wolf-Rayet stars}
 
  Next evolutionary state after LBV phase is the WR star of nitrogen sequence (WN). We consider WN state using two stars of WN8 subtype: FSZ35 which is located in M33 galaxy \citep{fsz35} and WR156 -- in our Galaxy \citep{meWR156}.   Performed modeling allows both to determine parameters of the stars as well as to refine the evolutionary state of GR\,290.

  \vspace{-5mm} 
\begin{table}\centering
\caption{Derived properties of studied stars. $R_{2/3}$ is radius where the Rosseland optical depth is equal to 2/3, $T_{\rm eff}$ is effective temperature at $R_{2/3}$, $\dot{M}_{cl}$ is mass loss rate, $X_H$ the mass fractions of hydrogen.} 
\label{tab:parmodel}
\bigskip
\begin{tabular}{ll lll c ccc}
\hline
\multicolumn{1}{c}{Star} &\multicolumn{1}{c}{$V$}   &\multicolumn{1}{c}{Sp.}   & $L_*,10^5$                  & $\dot{M}_{cl},10^{-5}$      &  $T_{\rm eff}$&$R_{2/3}$     & $V_{\infty}$& $X_H$ \\
 \multicolumn{1}{c}{~}   &\multicolumn{1}{c}{[mag]} &\multicolumn{1}{c}{type}  &\multicolumn{1}{c}{[$\Lsun$]}&[$\Msun \mbox{yr}^{-1}$]     &    [kK]       &[$\Rsun$]     &[$\kms$]     &      [\%]\\
\hline
Cyg\,OB2\,\#7            & 10.55                    & ${\rm O3If_*}$  & 10        &  0.15                        &  43.2     &    18        & 3250         &  50 \\
Cyg\,OB2\,\#11           & 10.03                    & ${\rm O5.5Ifc}$ & 6.5        & 0.17                        &  36.0     &    20.7        & 2200         & 82 \\
\\
GR\,290  (Jan   2005)    & 17.24                    & WN11h           & 12         &  4.2                        &  23.7     &    65        & 200         &  29.5 \\
GR\,290  (Aug   2014)    & 18.74                    & WN8h            & 5.3        &  1.7                        &  32.8     &    23        & 400         &  29.5 \\ 
\\                                                                    
WR156                    & 11.01                    & WN8h            & 6.5        &  1.5                        &  35     &    22        & 550         &  30 \\ 
FSZ35                    & 18.78                    & WN8h            & 6.5        &  2.4                        &  35.5     &   21.3        & 750         &  17\\ 
\hline
\\
\end{tabular}
\end{table}
 \vspace{-15mm}
 
\section{Discussion}   

In this work we studied different stars having similar initial masses in order to track the stellar evolution at different moments of stellar life. The comparison of stellar parameters raise the following questions:
\begin{itemize}
\item The star Cyg\,OB2\,\#7 has the evidences of non-spherical wind structrure -- how does this feature influence the consecutive evolution of the star?
\item In star Cyg\,OB2\,\#11 we see the nitrogen abundance anomaly -- how does it evolve, will it disappear on the LBV stage when the star changes significantly and loses its envelopes? and how will its remains manifest itself on the following WR stage? 
\end{itemize}

\acknowledgements The study was supported by RFBR grant No. 16-02-00148. 

\bibliography{Maryevabib}
\end{document}